\documentstyle[aps,psfig]{revtex}
\begin{document}
\twocolumn[\hsize\textwidth\columnwidth\hsize\csname @twocolumnfalse\endcsname
\draft
\title{Signatures of a staggered-flux phase in the $t$-$J$ model
with two holes on a 32-site lattice}
\author{P. W. Leung\cite{email}}
\address{Physics Department, Hong Kong University of Science 
and Technology,
Clear Water Bay, Hong Kong}
\date{\today}
\maketitle
\begin{abstract}
We study the relevance of the staggered-flux phase in the $t$-$J$
model using a system with two holes on a 32-site lattice
with periodic boundary conditions.
We find a staggered-flux pattern in the current-current correlation
in the lowest energy $d$-wave state where there
is mutual attraction between the holes. This staggered correlation
decays faster with distance when the hole binding becomes stronger.
This is in complete agreement with
a recent study by Ivanov, Lee and Wen (Phys. Rev. Lett. 84, 3958, (2000)) based
on the $SU(2)$ theory,
and strongly suggests that the staggered-flux
phase is a key ingredient in the $t$-$J$ model.
We further show that this staggered-flux pattern 
does not exist in a state where
the holes repel each other.
Correlations of the  chirality operator ${\bf S}_1\cdot({\bf
  S}_2\times{\bf S}_3)$ show that the staggered pattern of the
chirality is closely tied to the holes.

\end{abstract}

]

The $t$-$J$ model is one of the most important microscopic models
in the study of high temperature superconductivity.
It describes a doped CuO$_2$ plane as a system of  holes 
(or Zhang-Rice singlets \cite{zr88}) moving in a
spin background described by the antiferromagnetic Heisenberg model.
The Hamiltonian is
\begin{equation}
{\cal H} = -t\sum_{\langle ij\rangle\sigma}(\tilde{c}^\dagger_{i\sigma}
\tilde{c}_{j\sigma}+{\rm H.c.})+J\sum_{\langle ij\rangle} 
({\bf S}_i\cdot {\bf S}_j
-\frac{1}{4}n_in_j),
\label{hamiltonian}
\end{equation}
where $\tilde{c}^\dagger_{i\sigma}$ and $\tilde{c}_{i\sigma}$ are the 
projected fermion
operators, and
$n_i\equiv\tilde{c}_{i\sigma}^\dagger\tilde{c}_{i\sigma}$ 
is the fermion
number operator.
Understanding the properties of this model is a major challenge
in the theoretical study of high temperature superconductivity.
In spite of the simple form of ${\cal H}$, solving the $t$-$J$ model 
is non-trivial due to the strong interaction of 
the fermion objects.
Mean-field theory solutions to the model
often involve a fictitious statistical
flux \cite{fluxphase}. The flux pattern 
in these ``flux phases" can be uniform or
staggered. In particular, it has been proposed that
the staggered-flux phase in the $t$-$J$ model might lower 
the energy of the system
and became the ground state  at parameters
relevant to experimental systems \cite{staggeredflux}.
However, being an abstract mathematical quantity, it is difficult to
find a suitable signature for the flux. Consequently independent confirmation
of the existence of  flux phases by numerical or experimental studies 
is difficult.
A previous attempt to search for flux phases using exact diagonalization
on small clusters \cite{dagotto} has been either negative or inconclusive.

As pointed out by Ivanov, Lee and Wen \cite{ilw99}, in the case
of the doped $t$-$J$ model a signature
for the staggered-flux phase can be found
in the current-current correlation.
Using a Gutzwiller-projected
$d$-wave pairing wavefunction, they found such a pattern
in the current correlation, namely, the hole current
goes around the elementary
square plaquettes in the counterclockwise and clockwise
directions, suggestive of positive and negative
fluxes through the plaquettes in a staggered manner.
They explained this observation by showing that the 
Gutzwiller-projected $d$-wave pairing 
wavefunction
is equivalent to the $SU(2)$ projected staggered-flux wavefunction.
Although this result shows that the concept of a staggered-flux
phase is relevant in the projected $d$-wave pairing wavefunction,
it does not answer the question whether  this
phase exists in the $t$-$J$ model.
The origin of the staggered-flux pattern in their study is
in the staggered-flux phase of the $SU(2)$ projected wavefunction.
It will be very interesting to check whether such
a pattern in the current correlation exists using an  approach
that is independent of the mean-field theory.

Motivated by this result, we search for signatures for the
staggered-flux phase in the $t$-$J$ model using exact diagonalization.
This method is free from any analytical or numerical approximations,
and thus can serve as an independent test for the result
in Ref.~\cite{ilw99}. 
One serious drawback of exact diagonalization
is that it is subjected to finite-size effect. This is particularly
serious when we study correlation functions where the size
and boundary conditions can have significant effects.
To overcome the finite-size effect as much as possible, we
use the largest cluster on which the $t$-$J$ model is
currently solvable by
exact diagonalization, namely, the 32-site cluster with periodic
boundary conditions. In order to study the current correlation,
we need to dope the system with at least two holes.
Such a two-hole $t$-$J$ model on a 32-site cluster has
recently been solved using exact diagonalization \cite{clg98a,clg98b}.
In this study we use the same lowest energy
$d$-wave states at different $J/t$ as
in Ref.~\cite{clg98b}. They have $d_{x^2-y^2}$ symmetry and
 are the respective ground states when $J/t$ is not too small.
Our primary interest is in the case $J/t=0.3$ which is
believed to be an appropriate value to describe the doped
cuprates. With this parameter, the $d$-wave state 
is a weak two-hole 
bound state --- in the sense that it has a negative but small 
two-hole binding energy
$E_b/t=-0.05146$. The root-mean-square separation of the two holes
is $\sqrt{\langle r^2\rangle}=2.0587$.
Note that this value does not imply that the holes are tightly
bound. Due to the finite system size, two uncorrelated holes
on this cluster will have $\sqrt{\langle r^2\rangle}=2.3827$.

\begin{figure}
\centerline{
\psfig{figure=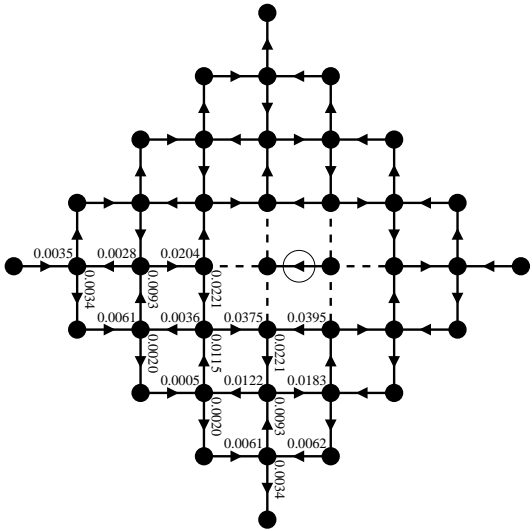,width=7cm}
}
\vspace{0.5cm}
\caption{The current correlation $\langle j_{kl}j_{mn}\rangle/x$
in the two-hole $t$-$J$ model at $J/t=0.3$.
$x\equiv1/16$ is the hole concentration.
The reference bond is indicated by an arrow with a circle.
Arrows on other bonds point along the directions of positive correlation.
Those arrows without numbers attached can be related by symmetry
to those that have.}
\label{fig:currenta}
\end{figure}

Once the desired wavefunction is found, it is straightforward
to evaluate the current correlation
$\langle j_{kl}j_{mn}\rangle$, where the current
on a bond linking the sites $k$ and $l$ is defined by
\begin{equation}
j_{kl}=it(\tilde{c}^\dagger_{k\sigma}\tilde{c}_{l\sigma}
-\tilde{c}^\dagger_{l\sigma}\tilde{c}_{k\sigma}).
\end{equation}
Following the notation of Ref.~\cite{ilw99}, we display the
current correlation divided by the hole concentration,
$\langle j_{kl}j_{mn}\rangle/x$.
The result for $J/t=0.3$ is shown in Fig.~\ref{fig:currenta}.
It is
to be compared with Fig.~1 of Ref.~\cite{ilw99}.
We note the very good agreement between these two results:
not only all (except one) directions of the correlation  agree,
but also the corresponding numbers agree to the same order of magnitude.
Given that the two results are obtained by very different
approaches, their agreement is very surprising. 
The only exception to the agreement in the direction of the
correlation is the one on
the opposite side of the same square plaquette as the reference
bond. This can be explained by a 
feature in the hole-hole correlation function,
\begin{equation}
C_{hh}(r)\equiv \langle (1-n_r)(1-n_0)\rangle,
\end{equation}
of the $d$-wave state that we use.
While $C_{hh}(1)>C_{hh}(\sqrt{2})$ in the projected $d$-wave 
pairing state \cite{i99},
the opposite is true in our $d$-wave state.
In fact the holes in our $d$-wave state
 have the largest probability to be at a 
distance $\sqrt{2}$ apart \cite{cofr}.
When the holes are at the opposite vertices of the same plaquette,
the hole current along the two opposite sides
of the plaquette should be along the same direction.

\begin{figure}
\centerline{
\psfig{figure=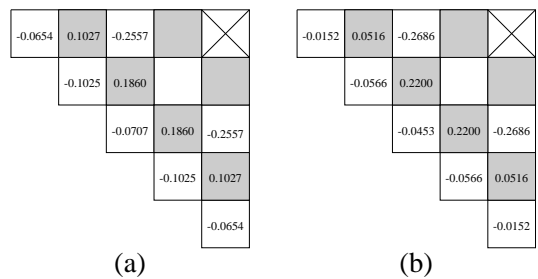,width=7cm}
}
\vspace{0.5cm}
\caption{The vorticity correlation $C_{VV}(r)$
at (a) $J/t=0.3$, and (b) $J/t=0.8$. The reference plaquette
at $r=0$ is indicated by a cross inside it.
Only part of the cluster
is shown.}
\label{fig:vorticity}
\end{figure}

In order to display the staggered pattern of the current
correlation more clearly, we follow Ref.~\cite{ilw99}
and define the vorticity $V({\bf r})$ of a square plaquette
centered  at {\bf r}
by summing up the currents around
it in the counterclockwise direction.
The vorticity-vorticity correlation divided by the hole
concentration, 
\begin{equation}
C_{VV}(r)\equiv\langle V(r)V(0)\rangle/x,
\end{equation}
is
shown in Fig.~\ref{fig:vorticity}(a).
Note that $C_{VV}$ can be computed from the current correlation, i.e.,
Fig.~\ref{fig:vorticity}(a) is just another way of presenting the
same data  in Figs.~\ref{fig:currenta}.
Fig.~\ref{fig:vorticity}(a) 
is to be compared to Fig.~1(b) of Ref.~\cite{ilw99}.
They all have the same staggered pattern.
Note that this staggered pattern has a $\pi$ phase shift, i.e.,
the sign of $\langle V(r)V(0)\rangle$ is $(-1)^{r_x+r_y+1}$.
This $\pi$ phase shift was explained within
the $SU(2)$ picture in Ref.~\cite{ilw99} by assuming the
pairing of holes of opposite vorticities, i.e., circulating
in opposite directions.

Since the hole current results from the motion of the holes,
it is intuitively clear that the correlation between the current
in different bonds 
must be related to the correlation between
the holes in the corresponding locations.
Therefore, in addition to the above weak bound state at $J/t=0.3$,
we also study the wavefunction with the same symmetry but
at $J/t=0.8$ where the holes 
are more tightly bound with $E_b/t=-0.44423$
and $\sqrt{\langle r^2\rangle}=1.5120$.
The purpose is to study how the current correlation changes
when the binding of the holes is stronger.
In Fig.~\ref{fig:vorticity}(b) we show the vorticity correlation at
$J/t=0.8$. 
We note that it has the same staggered with a $\pi$ phase shift pattern
as that at $J/t=0.3$.
Since the holes are more tightly bound in the present case,
we expect the current correlation, and therefore the vorticity
correlation, to have a shorter
range. This is indeed the case when we compare it with 
Fig.~\ref{fig:vorticity}(a). $C_{VV}(r)$ decreases
much faster in the present case. 

\begin{figure}
\centerline{
\psfig{figure=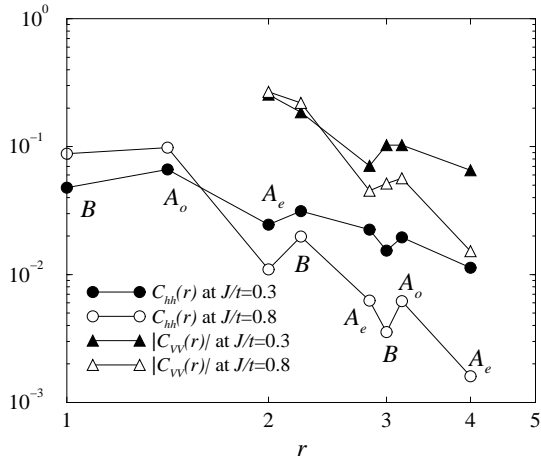,width=7cm}
}
\vspace{0.5cm}
\caption{The hole correlation $C_{hh}(r)$ and the vorticity
correlation $|C_{VV}(r)|$ of the $d$-wave states at
$J/t=0.3$ and 0.8.
$A_o$ ($A_e$) indicates that the pair of holes at 0 and $r$
are on the same sublattice and $r$ is an odd (even) site.
$B$ indicates that they are on opposite sublattices.
}
\label{fig:scale}
\end{figure}

In Ref.~\cite{ilw99} it was observed that 
in the projected $d$-wave pairing state the hole-hole
and vorticity-vorticity correlations decay with the same power law.
In Fig.~\ref{fig:scale} we plot these two quantities in 
our $d$-wave states at  $J/t=0.3$ and $0.8$. 
As noted before \cite{ilw99,i99}, $C_{hh}(r)$ depends on
whether 0 and $r$ are on the same sublattice, and if so, 
whether $r$ is an even ($r_x$ and $r_y$
are even) or odd ($r_x$ and $r_y$ are odd) site.
Those on the same sublattice and odd sites are enhanced, while those
on the same sublattice and even sites are suppressed relative
to those on different sublattices. 
If we assume that the hole correlation decays with
a power law, $C_{hh}(r)\approx 1/r^\alpha$, then the best fit
gives $\alpha\approx 1.15<2$ at $J/t=0.3$, making it an unbound state
when the system size goes to infinity.
Being a tightly bound state, the hole correlation
at $J/t=0.8$ decays much faster than that at $J/t=0.3$ ---
the corresponding $\alpha\approx  3.04$ is significantly 
larger than 2.
The same trend is found for the vorticity correlation.
Nevertheless, with the limited number of data points
we are not able to say conclusively whether they decay with the same
law, nor whether they decay with  power laws at all.

\begin{figure}
\centerline{
\psfig{figure=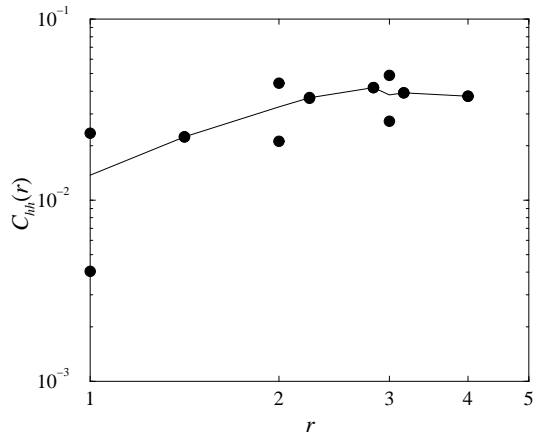,width=7cm}
}
\vspace{0.5cm}
\caption{The hole correlation $C_{hh}(r)$ of the $p_y$-wave state
at $J/t=0.3$. Due to the lack of the rotational symmetry $C_4$
(i.e. a rotation by $\pi/2$)
in the wavefunction, there may be two inequivalent points at the
same $r$. Their values are indicated by filled circles. The solid
line joins their mean values.}
\label{fig:phole}
\end{figure}

\begin{figure}
\centerline{
\psfig{figure=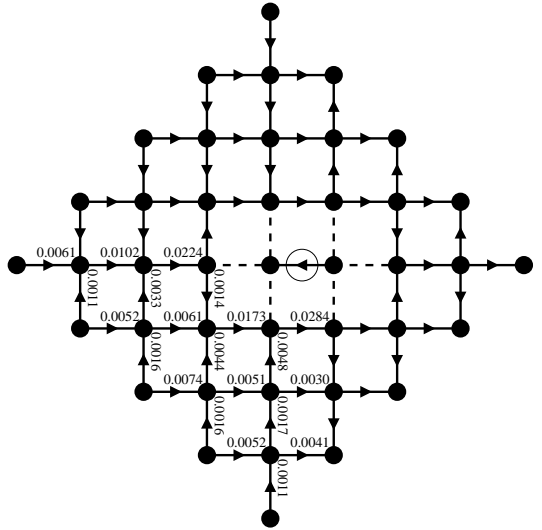,width=7cm}
}
\vspace{0.5cm}
\caption{Same as Fig.~\ref{fig:currenta} but for the $p_y$-wave state.
Due to the lack of the rotational symmetry $C_4$, there is an
inequivalent correlation where the reference bond is a vertical one.
It is not shown for the sake of simplicity.
}
\label{fig:pcurrent}
\end{figure}

\begin{figure}
\centerline{
\psfig{figure=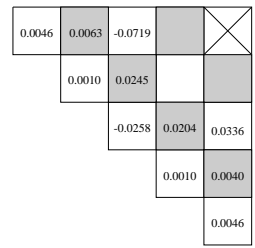,width=3.2cm}
}
\vspace{0.5cm}
\caption{The vorticity correlation $C_{VV}(r)$ of the $p_y$-wave
  state. }
\label{fig:pvorticity}
\end{figure}

The above results show that a staggered-flux pattern exists
in our exact $d$-wave state where the hole correlation decays with distance.
In Ref.~\cite{ilw99} such a pattern is interpreted as the binding
of holes of opposite vorticities in the $SU(2)$ language.
It is therefore interesting to see, using our exact 
numerical results, if such a pattern
exists in a state where the holes repel.
For this purpose we use a low-lying 
excited state of the same two-hole system
at $J/t=0.3$. It is a doubly degenerate 
$p$-wave state with
total momentum $(\pi,\pi)$ and is a spin singlet \cite{pwave}. 
Fig.~\ref{fig:phole} shows the hole correlation
of the $p_y$-wave state.
In contrary to the $d$-wave ground state,
the holes in this state repel each other ---
$C_{hh}$ increases with $r$ for small $r$. 
The mean-square separation between the holes is 
$\sqrt{\langle r^2\rangle}=2.5376$.
Fig.~\ref{fig:pcurrent} shows the current correlations in the
$p_y$-wave state.
We observe that the magnitudes of the correlations are small
even at small distances, and there is definitely no staggered-flux pattern.
The same data plotted as vorticity correlation in Fig.~\ref{fig:pvorticity}
again confirms these observations.
This is consistent with the interpretation that the staggered-flux
pattern results from the pairing of the holes.


\begin{figure}
\centerline{
\psfig{figure=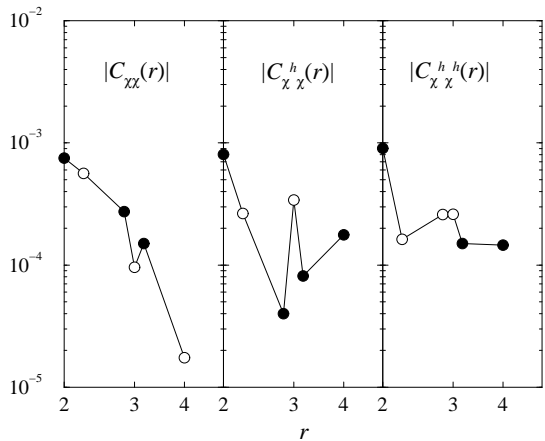,width=7cm}
}
\vspace{0.5cm}
\caption{Correlations of the chirality operators of the $d$-wave state
  at $J/t=0.3$ in the same logarithmic scale. 
Empty and solid circles represent positive and
  negative values respectively. }
\label{fig:chi}
\end{figure}

We now return to the $d$-wave state at $J/t=0.3$ and search for
another signature of the staggered-flux phase.
It has been suggested that a physical interpretation of the
statistical flux in the mean-field theory
can be found in the spin chirality operator \cite{ln92},
\begin{equation}
\chi_4=n_4\,{\bf S}_1\cdot({\bf S}_2\times{\bf S}_3),
\end{equation}
where 1, 2, 3, 4 are the vertices of a square plaquette in the
counterclockwise direction.
In Fig.~\ref{fig:chi} we plot the correlation
$C_{\chi\chi}(r)\equiv\langle\chi_r\chi_0\rangle$ for the $d$-wave
state at $J/t=0.3$. It is obvious that this
correlation exhibits a short-range staggered pattern with a $\pi$ phase shift,
but it decays very  rapidly  with distance. Since the chirality is
related to a magnetic flux, which may in turn affect the movement of
the holes, it is natural to associate the chiral order to the holes.
To demonstrate the
connection between the chirality and the hole current, 
we follow Ref.~\cite{sil00} and define the spin chirality
operator around a hole,
\begin{equation}
\chi_4^h=(1-n_4)\,{\bf S}_1\cdot({\bf S}_2\times{\bf S}_3).
\end{equation}
The chiral order set up by a hole is displayed in the correlation
$C_{\chi^h\chi}(r)\equiv\langle \chi_r\chi^h_0\rangle$ in Fig.~\ref{fig:chi}.
It has a staggered pattern with a $\pi$ phase shift. Compared to
$C_{\chi\chi}(r)$, it decays slower, but has strong fluctuation as a
function of $r$. Finally,
we plot the correlation between the chirality around the two holes,
$C_{\chi^h\chi^h}(r)\equiv\langle \chi^h_r\chi^h_0\rangle$.
Its decay with distance is again much slower than in the case of
$C_{\chi\chi}(r)$. 
These correlations show that the chiral order must
be tied to the holes. Unfortunately, with the limited data we are not
able to deduce conclusively whether long-range chiral order exists in
the current system. We also note that similar results on the chirality
correlations were obtained with the projected $d$-wave pairing
wavefunction \cite{sil00}.

In conclusion, we have looked for signatures for the staggered-flux
phase in the $t$-$J$ model by calculating the current
and chirality correlations
in the lowest-energy $d$-wave state of the two-hole $t$-$J$
model on a 32-site cluster with periodic
boundary conditions.
The current correlation clearly shows 
a staggered-flux pattern. Note that in the
$d$-wave states at different $J/t$, the holes
are mutually attractive --- $C_{hh}(r)$ decays with distance.
In this case the staggered correlation in the vorticity
decays faster when the attraction between the holes becomes stronger.
In the $p$-wave state where the holes repel ($C_{hh}(r)$ increases
with distance at small distances), no such staggered-flux pattern
is found. Chirality correlations also show a similar staggered
pattern. Although the numbers are small, they show that any long-range
pattern in the chirality correlation must be related to the holes.
Our results on the current correlation are consistent with
a recent $SU(2)$ study by Ivanov, Lee and Wen \cite{ilw99}. 
Being a completely independent approach,
the fact that our unbiased numerical result agrees with theirs
gives strong support to the notion that the staggered-flux
phase is a key ingredient of the ground state in the $t$-$J$ model,
at least in the low doping regime.


We are grateful to P. A. Lee for pointing out their work (Ref.~\cite{ilw99})
to us and for helpful discussions. This work was supported by the RGC
of Hong Kong under grant number HKUST6144/97P.

%
%

%
%


\begin{references}
\bibitem{pwave} Details of this $p$-wave state will be
published elsewhere.
\end{references}
\end{document}